\documentclass[superscriptaddress,twocolumn,english,10pt,prl,showpacs]{revtex4-1}
\usepackage[latin9]{inputenc}
\usepackage[T1]{fontenc}
\usepackage{amsmath}
\usepackage{amssymb}
\usepackage{graphicx}
\usepackage{babel}
\usepackage{float}
\usepackage[colorlinks]{hyperref}
\usepackage{color}

\definecolor{applegreen}{rgb}{0.55, 0.71, 0.0}

\global\long\def\ket#1{\left| #1\right\rangle }
\global\long\def\bra#1{\left\langle #1 \right|}

\global\long\def\av#1{\left\langle #1 \right\rangle }
\global\long\def\tr{\text{tr}}

\global\long\def\pd{\partial}
\global\long\def\im{\text{Im}}

\global\long\def\sgn{\text{sgn}}

\global\long\def\abs#1{\left|#1\right|}

\global\long\def\bs#1{\boldsymbol{#1}}
\global\long\def\t#1{\text{#1}}

\begin{document}

\title{Mixed-order symmetry-breaking quantum phase transition far from equilibrium}

\author{T. O. Puel}
\email{tharnier@csrc.ac.cn}
\address{Beijing Computational Science Research Center, Beijing 100193, China}
\affiliation{CeFEMA, Instituto Superior T\'{e}cnico, Universidade de Lisboa Av. Rovisco
Pais, 1049-001 Lisboa, Portugal}
\affiliation{Zhejiang Institute of Modern Physics, Zhejiang University, Hangzhou,
Zhejiang 310027, China}
\affiliation{Zhejiang Province Key Laboratory of Quantum Technology and Devices, Zhejiang University, Hangzhou 310027, China}

\author{Stefano Chesi}
\email{stefano.chesi@csrc.ac.cn}
\affiliation{Beijing Computational Science Research Center, Beijing 100193, China}
\affiliation{Department of Physics, Beijing Normal University, Beijing 100875, China}

\author{S. Kirchner}
\email{stefan.kirchner@correlated-matter.com}
\affiliation{Zhejiang Institute of Modern Physics, Zhejiang University, Hangzhou,
Zhejiang 310027, China}
\address{Zhejiang Province Key Laboratory of Quantum Technology and Devices, Zhejiang University, Hangzhou 310027, China}

\author{P. Ribeiro}
\email{pedrojgribeiro@tecnico.ulisboa.pt}
\affiliation{CeFEMA, Instituto Superior T\'{e}cnico, Universidade de Lisboa Av. Rovisco
Pais, 1049-001 Lisboa, Portugal}
\affiliation{Beijing Computational Science Research Center, Beijing 100193, China}

\date{\today}

\begin{abstract}
We study the current-carrying steady-state of a transverse field Ising chain coupled to magnetic thermal reservoirs and obtain the non-equilibrium phase diagram as a function of the magnetization potential of the reservoirs. 
Upon increasing the magnetization bias we observe a discontinuous jump of the magnetic order parameter that
coincides with a divergence of the correlation length. 
For steady-states with a non-vanishing conductance, the entanglement entropy at zero temperature displays a bias dependent logarithmic correction that violates the area law and differs from the well-known equilibrium case. 
Our findings show that out-of-equilibrium conditions allow for novel critical phenomena not possible at equilibrium.
\end{abstract}

\maketitle

\paragraph{Introduction:}
Non-equilibrium phases of quantum matter in open systems is a topical issue of immediate experimental relevance 
\cite{Pothier1997,Anthore.03,Chen.09,Brantut2013,Fink2017, Fitzpatrick2017}. However, a theoretical framework for the description of out-of-equilibrium strongly-correlated systems is at present incomplete and requires the further development of reliable techniques for non-equilibrium conditions (see, e.g., Refs.~\cite{Sieberer2015,Jin2016,Kshetrimayum2017} 
and references therein). The influence of a non-thermal drive on phase boundaries and quantum critical points (QCP) is of particular interest.

An important class of non-equilibrium states are current-carrying steady-states (CCSS) that emerge in the long-time limit of systems coupled to  reservoirs which are held at different thermodynamic potentials.
These states are characterized by a steady flow of otherwise conserved
quantities, such as energy, spin or charge. They can be realized in solid-state
devices \cite{Pothier1997,Anthore.03,Chen.09} and have recently also became available in cold atomic setups \cite{Brantut2013}.

For Markovian processes, substantial progress has been made due  the discovery of exact solutions for boundary driven Lindblad dynamics \cite{Prosen2008, Prosen2010a, Prosen2011a, Prosen2014} allowing for the characterization of certain non-equilibrium phases and phase transitions.  
In these cases, however,  the Markovian condition substantially simplifies the dynamics. As a result, its validity is confined to extreme non-equilibrium conditions (e.g., large bias) that cannot be connected to thermal equilibrium \cite{Ribeiro2014e, Ribeiro2015f}.  
Non-thermal steady-states in Luttinger liquids have also been studied
\cite{Gutman.2008,Gutman.2009, Dinh.2010}, but the results are less general than their equilibrium counterparts.
Other methods to study CCSS include, 
looking at the asymptotic dynamics in pairs of semi-infinite quantum wires following quenches of the hopping connecting the pairs
\cite{Lancaster2010, Lancaster.2011, Sabetta2013, Bernard2012, Calabrese2008},
Bethe ansatz-based approaches \cite{Mehta.2006,Caux2016} that exploit the properties of integrable systems, hybrid approaches involving Lindblad dynamics \cite{Dorda.14} and more phenomenological approximations based on Boltzmann kinetic equations \cite{Lunde.2007,Lunde.2009}.

Another guiding element is the occurrence of scaling and criticality, which signal the absence of intrinsic energy scales and make the system particularly susceptible to any non-equilibrium drive \cite{Kirchner2005, Mitra.06, Takei.08, Kirchner.09, Chung2009,Chung2012, Aoki.14}. 
Phase-transitions under non-equilibrium conditions \cite{Ribeiro2013b, Lesanovsky2013,Horstmann2013,Genway2014,Ribeiro.15,Zamani.2016,Sieberer2015,Znidaric2015, Marino.16, Hannukainen2018}
were shown to allow intrinsic non-equilibrium universal properties, not seen at  equilibrium. Nevertheless, a systematic approach describing CCSS is not available and exact solutions therefore must serve as a guiding principle.

\begin{figure}
\centering{}\includegraphics[width=0.99\columnwidth]{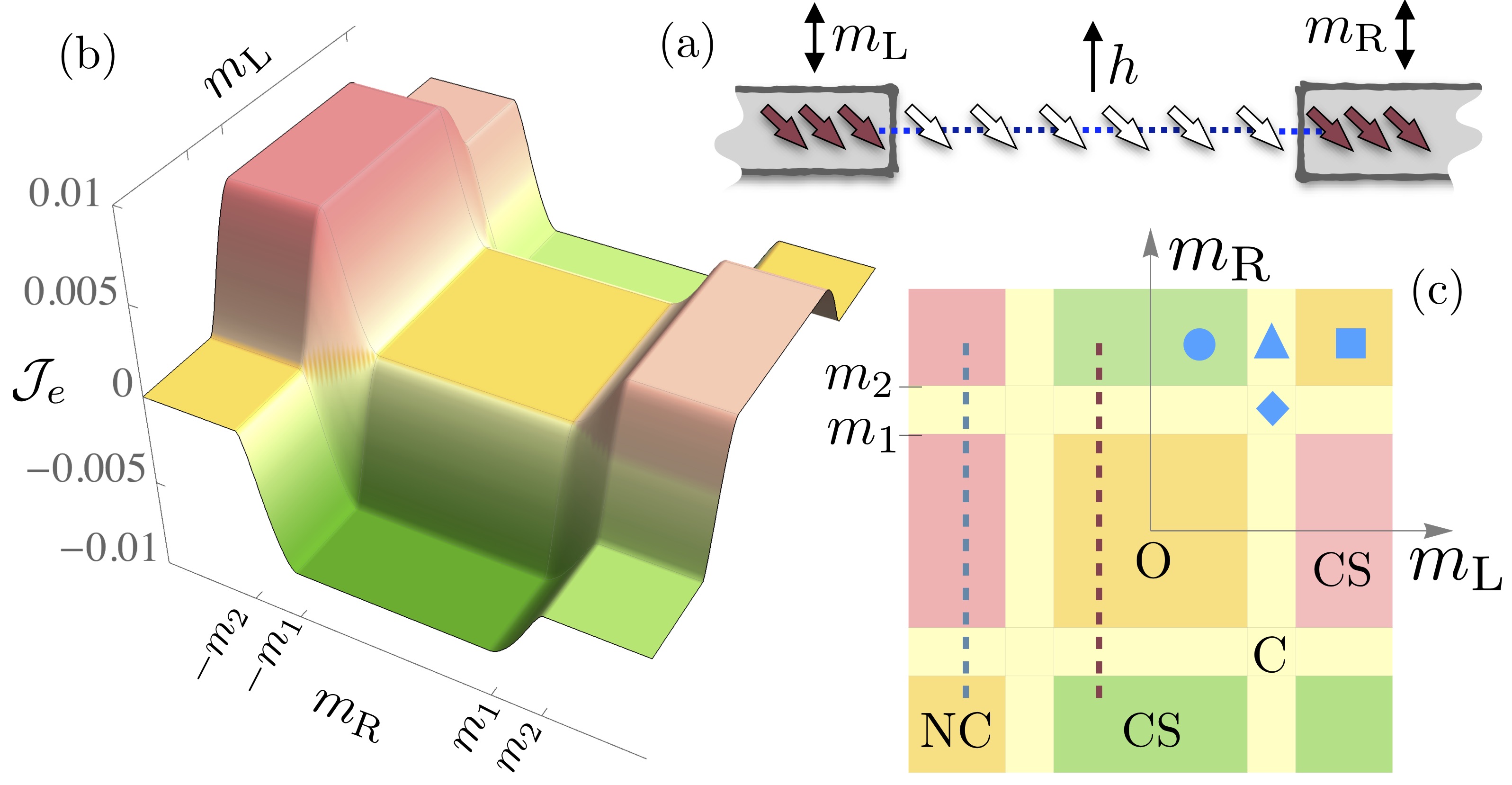}\caption{\label{fig: phaseDiagram} 
(a) Sketch of the model - transverse field Ising chain coupled at its edges to magnetic reservoirs, $\text{L}$ and $\text{R}$, held at magnetizations $m_\text{L}$ and $m_\text{R}$ respectively. 
(b) Energy current, ${\cal J}_e$, flowing through the chain as function of  $m_{\text{L}}$ and $m_{\text{R}}$.
(c) Schematic phase diagram - color coding matches that of (b); 
The phase labels are: O for ordered, NC for non-conducting, C for conducting, and CS for conducting saturated. The properties of these phases are discussed in the text. Properties displayed in Figs.~\ref{fig: order parameter} and \ref{fig:currentEntropy} correspond to the parameters along the dashed lines; geometric symbols mark the parameters  used in Fig.~\ref{fig: occupation number}. Here $\Gamma_{L,R}=0.01$.}
\end{figure}

In this letter we discuss an order-disorder symmetry breaking transition induced by  non-equilibrium conditions in one of such exactly-solvable models, i.e.,  a spin chain that admits an exact solution by a mapping to a non-interacting fermionic system. Besides presenting the phase diagram and a characterization of various non-equilibrium phases, we identify a remarkable mixed-order quantum phase transition, where a discontinuous jump of the order parameter occurs in the presence of a divergent correlation length. The coexistence of such defining features of first- and second-order phase transitions implies the emergence a universality class specific to non-equilibrium conditions, for which an effective field-theoretic description is yet to be developed.

\begin{figure}
\centering{}\includegraphics[width=0.99\columnwidth]{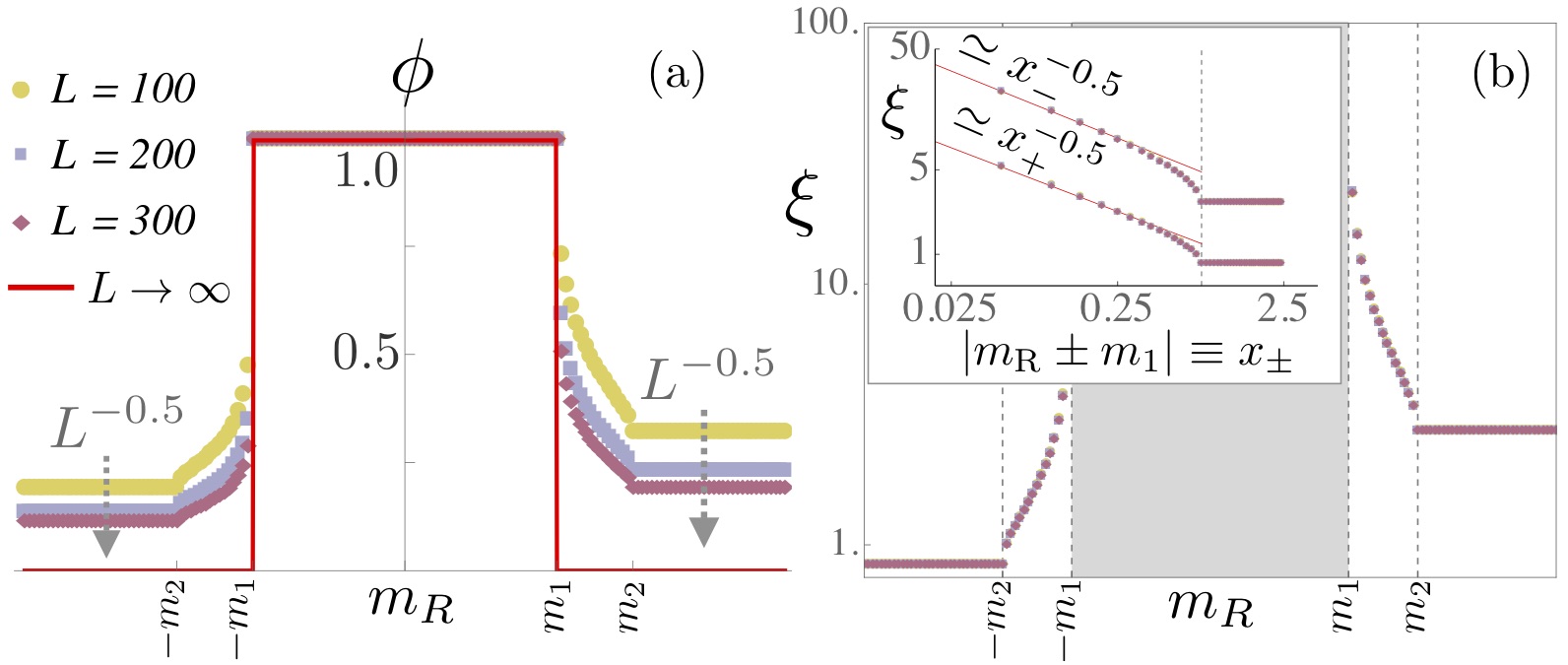}\caption{\label{fig: order parameter} 
(a) Order parameter, $\phi$, computed for parameters along the red-dashed line in Fig.\ref{fig: phaseDiagram}-(c) for different system sizes. 
(b) Correlation length, $\xi$, for parameters along the red-dashed line in Fig.\ref{fig: phaseDiagram}-(c) and for the same system sizes in panel (a).  The inset shows the log-scaling of $\xi$ near the transition points $m_\text{R}=\pm m_{1}$. }
\end{figure}

\paragraph{Model:}
The model we consider is depicted in 
 Fig.~\ref{fig: phaseDiagram}-(a) and consists of an Ising spin chain of length $L$, exchange coupling $J$ and an applied transverse field $h$, coupled to two zero-temperature magnetic reservoirs at 
$r=r_{\text{L}} \equiv 1$ and $r=r_{\text{R}} \equiv L$ respectively.
The total Hamiltonian is given by 
\begin{equation}
\label{eq:Hamiltonian}
H=-J\sum_{r=1}^{L-1}\sigma_{r}^{x}\sigma_{r+1}^{x}-h\sum_{r=1}^{L}\sigma_{r}^{z}+\sum_{l=\text{L},\text{R}}\left(H_{l}+H_{\text{C},l}\right),
\end{equation}
where $\sigma_{r}^{x,y,z}$ are the
Pauli matrices acting on site $r$. The reservoirs are described by isotropic XY models, ${ H_{l}=-J_{l}\sum_{r\in\Omega_{l}}\left(\sigma_{r}^{x}\sigma_{r+1}^{x}+\sigma_{r}^{y}\sigma_{r+1}^{y}\right)-m_{l}M_l }$
 with $\Omega_{\text{L}}=\left\{-\infty,...,0\right\}$, $\Omega_{\text{R}}=\left\{L+1,...,\infty\right\}$,
 and  the magnetization $M_{l}=\sum_{r\in \Omega_{l}}\sigma_{r}^{z}$ (which is a good quantum number, i.e. $\left[H_{l},M_{l}\right]=0$).
The chain-reservoirs coupling Hamiltonians are
${ H_{\text{C},l}=-J'_{l}\left(\sigma_{r'_{l}}^{x}\sigma_{r_{l}}^{x}+\sigma_{r'_{l}}^{y}\sigma_{r_{l}}^{y}\right) }$,
with ${ r'_{\text{L}}=0 }$ and ${ r'_{\text{R}}=L+1 }$. 
Each reservoir is characterized by a set of gapless magnetic excitations within an energy bandwidth $J_{l}$ and  the average value of $M_l$ is set by the magnetic potential $m_{l}$. Below we use $J$ as our unit of energy, i.e. $J=1$.

\paragraph{Non-equilibrium order-disorder phase transition:}
The ground-state of
the chain Hamiltonian $H_\text{C}$ [the first two terms of Eq.~\eqref{eq:Hamiltonian}] has a continuous phase transition for $h=\pm1$  that separates a $\mathbb{Z}_2$ symmetry broken state from a paramagnetic one. 
The symmetry-broken state can be characterized by an order parameter  
$\phi=\lim_{h_{x}\to0}\lim_{L\to\infty}\left\langle \sigma_{r}^{x}\right\rangle, \forall \, r $, with $h_{x}$ a magnetic field along $x$ that explicitly breaks the $\mathbb{Z}_{2}$ symmetry.
$\phi$ vanishes as $\abs{\phi}=\left(1-h^{2}\right)^{1/8}$ \cite{Barouch.71}
as the transition point is approached from the ordered side, {\itshape i.e.} $|h| \rightarrow 1$, with the critical exponent $\beta=1/8$. 
The correlation length diverges as $\xi \propto \left(1-h^{2}\right)^{-\nu}$ with $\nu=1$. 
This phase transition is in the universality class of the 2d classical Ising model and thus the QCP is described by a  $\phi^4$ theory. 

Our primary concern in this letter is
the steady-state phase diagram that emerges far from equilibrium when  $J'_l\neq 0$.
The energy drained from the left reservoir is  $\mathcal{J}_{e}=-i\left\langle \left[H,H_{\t L}\right]\right\rangle $, which equals the steady-state energy current in any cross section along the chain (detailed calculations are provided in the next section). 
The current $\mathcal{J}_{e}$ is depicted in Fig.~\ref{fig: phaseDiagram}-(b) as a function of the left and right magnetic potentials,
while Fig.~\ref{fig: phaseDiagram}-(c) schematically shows its corresponding non-equilibrium phase diagram. 
We consider the case $\left|h\right|<1$, for which the equilibrium phase is ordered.
Interestingly, the ordered state survives a non-vanishing coupling to the reservoirs for $\abs{m_{\text{L,R}}}<m_1$, with $m_1=2\left(-h+1\right) > 0$. 
The order parameter
along the dashed-red segment of Fig.~\ref{fig: phaseDiagram}-(c) is depicted in Fig.~\ref{fig: order parameter}-(a). Within the ordered phase  $\phi$ does not depend on $m_\text{R}$. 
 At $|m_\text{R}| = m_1$, $\phi$ drops discontinuously to zero as $L\to\infty$
, and this limit is  approached as  $\phi \sim L^{-1/2}$
in the disordered phase ($|m_\text{R}| > m_1$). In this region we have also computed the correlation length $\xi$, shown in Fig.~\ref{fig: order parameter}-(b). For $m_\text{R} \rightarrow \mp m_1$ from the disordered phase we find a divergent behavior $\xi\propto \abs{m_\text{R}\pm m_1}^{-\lambda} $, compatible with a critical exponent $\lambda=1/2$.
\footnote{As in equilibrium, we expect the correlation length to diverge also on the ordered side, however, a confirmation is beyond the current approach.}
 Our results imply that the discontinuous vanishing of $\phi$ at $|m_\text{R}|= m_1$ in the $L\rightarrow \infty$ limit, 
 a characteristic feature of a first-order phase transition,
 is accompanied by a divergent correlation length, 
 a hallmark of continuous phase transitions.
Therefore, such a behaviour  cannot be accommodated within an equilibrium effective description. Below, some immediate implications of this significant finding will be further substantiated and analyzed. In particular, we will present the order-disorder transition in the context of a detailed description of the model and its other interesting non-equilibrium properties.


\paragraph{Methodology:}
The full Hamiltonian, $H$, can be represented in terms of fermions through the so-called Jordan-Wigner mapping \cite{Lieb1961}, $\sigma_{r}^{+}=e^{i\pi\sum_{r'=0}^{r-1}c_{r'}^{\dagger}c_{r'}}c_{r}^{\dagger}$,
where $c_{r}^{\dagger}/c_{r}$ creates/annihilates a spinless fermion
at site $r$. This leads to a Kitaev chain~\cite{Kitaev2001,comment1} in contact with two metallic reservoirs at chemical potentials $\mu_{\text{L},\text{R}} =2 m_{\text{L},\text{R}}$. The topological non-trivial phase corresponds to the ordered phase of the original spin model. The transformed Hamiltonian is quadratic and the chain contribution is given by $H_{\text{C}}=\frac{1}{2}\boldsymbol{\Psi}^{\dagger}\bs H_{\t C}\boldsymbol{\Psi}$,
with 
$\boldsymbol{\Psi}^{\dagger}=\left(c_{1}^{\dagger},\ldots,c_{L}^{\dagger},c_{1},\ldots,c_{L}\right)$,
and where $\bs H_{\text{C}}$ is a $2L\times2L$ Hermitian matrix respecting
the particle-hole symmetry condition $\bs S^{-1}\boldsymbol{H}_{\t C}^{T}\bs S=-\boldsymbol{H}_{\t C}$
with $\bs S=\tau^{x}\otimes\bs 1_{L\times L}$ and where $\tau^{x}$
interchanges particle and hole subspaces. 
In the fermionic representation,
any correlation function can be described in terms of the retarded, advanced and 
and Keldysh components
of the single-particle Green's function \cite{SM}. 
 
In the following we make the simplifying
assumption that the bandwidths of the reservoirs, $J_{l=\text{L},\text{R}}$, are much
larger than all other energy scales (``wide band limit'').
In this limit, the coupling to each reservoir $l$ is completely determined by $\Gamma_{l}=\pi {J'_l}^{2} D_{l}$, the hybridization energy scale, with $D_{l}$ being the local density of states of the reservoir.
Furthermore, we can define the non-Hermitian single-particle operator 
$\bs K=\bs H_{\t C}-i\sum_{l=\text{L,R}}\left(\bs{\gamma}_{l}+\hat{\bs{\gamma}}_{l}\right)$, 
with $\bs{\gamma}_{l}=\Gamma_{l}\ket{r_{l}}\bra{r_{l}}$
and $\hat{\bs{\gamma}}_{l}=\Gamma_{l}\ket{\hat{r}_{l}}\bra{\hat{r}_{l}}$, and
where $\ket r$ and $\ket{\hat{r}}=\bs S\ket r$ are single-particle states.
We assume that $\bs K$ is diagonalizable, having 
right and left eigenvectors $\ket{\alpha}$ and $\bra{\tilde{\alpha}}$, with associated eigenvalues $\lambda_{\alpha}$.

Equal-time observables can be obtained from the single-particle density
matrix defined as $\boldsymbol{\chi} \equiv \langle \boldsymbol{\Psi}\boldsymbol{\Psi}^{\dagger}\rangle$,
which is explicitly given by
\begin{multline}
\bs{\chi}=\frac{1}{2}+\sum_{l=\t L,\t R}\sum_{\alpha\beta}\ket{\alpha}\bra{\beta}\times\\
\langle\tilde{\alpha}|\left[\bs{\gamma}_{l}I_{l}\left(\lambda_{\alpha},\lambda_{\beta}^{*}\right)-\hat{\bs{\gamma}}_{l}I_{l}\left(-\lambda_{\alpha},-\lambda_{\beta}^{*}\right)\right]|\tilde{\beta}\rangle\label{eq:chi}
\end{multline}
where 
$I_{l}\left(z,z'\right)=-\frac{1}{\pi}\frac{g\left(z-2m_{l}\right)-g\left(z'-2m_{l}\right)}{z-z'}$
with $g\left(z\right)=\ln\left(-i\sgn\left[\im\left(z\right)\right]z\right)$.

\begin{figure}[t]
\includegraphics[width=0.99\columnwidth]{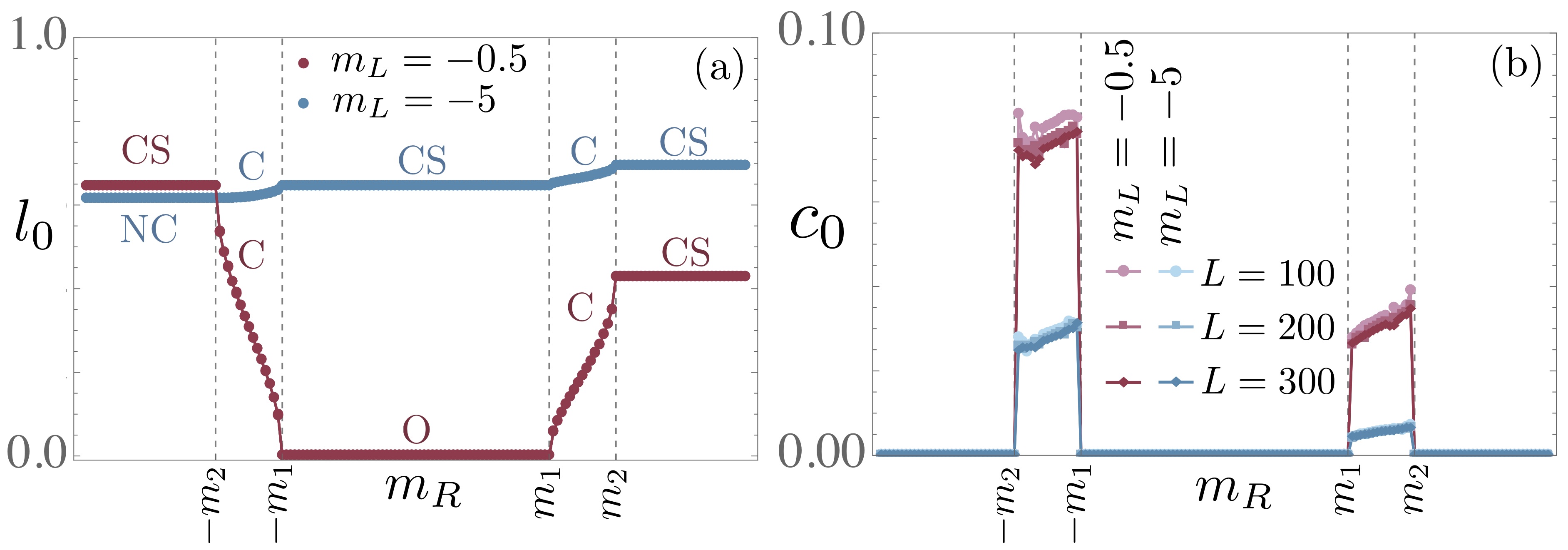}
\caption{
Scaling analysis of the entanglement entropy, $E_{\t S_\ell} \simeq l_0 \ell+ c_0 \log \ell + c_1$, of a sub-system ${\t S_\ell}$.  Red (blue) data points correspond to parameters along the red-dashed (blue-dashed) line in Fig.\ref{fig: phaseDiagram}-(c). 
The color coding in panel (b) shows data points for different values of $L$. 
}
\label{fig:currentEntropy}
\end{figure}

The current of energy which drains from the left reservoir is equal to the steady-state energy current in any cross section along the chain,  thus can be obtained from  $\bs{\chi}$ as
$\mathcal{J}_{e} = - \frac{1}{2} \, \text{Tr} \left[ \boldsymbol{J}_{r}  \boldsymbol{\chi} \right]$, where $r$ is arbitrary and 
\begin{multline}
\boldsymbol{J}_{r} = -2ihJ \left[  (1+{\bs S})\ket{r-1}\bra{r} (1+{\bs S}) - \text{H.c.} \right] \, .
\end{multline}
The linear and non-linear thermal conductivity,  as well as other thermoelectric properties of the chain, are determined by $\mathcal{J}_{e}$. 

\paragraph{Results:}

As anticipated, $\mathcal{J}_{e}$ is able to discriminate between  different phases. We have shown in Fig.~\ref{fig: phaseDiagram}-(b) an example for $h=0.2$, illustrating the typical behavior and leading to the phase diagram sketched in Fig.~\ref{fig: phaseDiagram}-(c). 
Two phases with $\mathcal{J}_{e}=0$, NC and O, arise around the condition $m_\text{L}=m_\text{R}$.
Note, however, that this condition does not correspond to equilibrium for the fermionic system away from $m_{R}=m_{L}=0$. This is due to the fact that the non-interacting p-wave superconductor does not conserve the number of particles which in the spin representation translates to the non-conservation of the total magnetization.
A conducting phase, C, characterized by a non-zero conductance, $\pd_{m_{l={\t L},{\t R}}}\mathcal{J}_{e}\neq0$, arises for $\abs{ m_\text{L}}$ or $\abs{m_\text{R}}\in (m_1, m_2)$,
where $m_{2}>0$ is defined as $m_{2}=2(h+1)$.
A set of phases to which we refer as current-saturated, or CS, arise for $\abs{m_\text{L}}$ or $\abs{m_\text{R}}>m_2$ and are characterized by a finite current, $\mathcal{J}_{e}\neq 0$, and a vanishing conductance $\pd_{m_{l={\t L},{\t R}}}\mathcal{J}_{e} = 0$. 

In order to study the onset of order under non-equilibirum conditions, we have extended the equilibrium expression of the correlation function \citep{Lieb1961} to the general non-equilibrium case \cite{SM}.
In particular, the two-point correlation function,
$\mathbb{C}_{r,r'}^{\alpha\beta}=\langle \sigma_{r}^{\alpha}\sigma_{r'}^{\beta}\rangle -\langle \sigma_{r}^{\alpha}\rangle \langle \sigma_{r'}^{\beta}\rangle $, for $\alpha=\beta=x$ can be found in terms of $\bs{\chi}$ as follows:
\begin{eqnarray}
\mathbb{C}_{r,r'}^{xx} & = & \det\left[i\left(2\bs{\chi}_{\left[r,r'\right]}-1\right)\right]^{\frac{1}{2}},
\label{eq:C_xx}
\end{eqnarray}
where, for $r>r'+1$, $\bs{\chi}_{\left[r,r'\right]}$ is a $2\left(r-r'\right)$ 
matrix obtained as the restriction of $\bs{\chi}$ to the subspace
in which $\mathbb{P}_{rr'}^{T}=\sum_{u=r'+1}^{r-1}\left(\ket u\bra u+\ket{\hat{u}}\bra{\hat{u}}\right)+\ket{r_{+}}\bra{r_{+}}+\ket{r'_{-}}\bra{r'_{-}}$ acts as the identity, 
with $\ket{r_{\pm}}=\left(\ket r\pm\ket{\hat{r}}\right)/\sqrt{2}$. The full derivation of Eq.~\eqref{eq:C_xx} is given in \cite{SM}. 

Except for $\mathbb{C}_{r,r'}^{xx}$ in the ordered phase, O, all the other components of $\mathbb{C}_{r,r'}^{\alpha \beta}$, for $\alpha,\beta=x,y$,  decay exponentially. $\xi$ in Fig.~\ref{fig: order parameter}-(b) was obtained by fitting an exponentially decaying   $\mathbb{C}_{r,r'}^{xx}\propto e^{-\abs{r-r'}/\xi}$ to the numerical data generated by Eq.~\eqref{eq:C_xx}. 
For a finite system with $h_x=0$, since the $\mathbb{Z}_{2}$ symmetry is never broken,
$\phi$ can be computed by the relation $\phi^{2}=\lim_{L\to\infty}\mathbb{C}_{uL,u'L}^{xx}$, with $u,u'\in\left(0,1\right)$.
$\phi$ in Fig.~\ref{fig: order parameter}-(a) was computed in this way.  
Whenever $m_\text{R}$ or $m_\text{L}$ approaches the boundary $m_1$ of the ordered phase, we find that $\lambda(h)=1/2$ for $0<h<1$, except for $h=1/2$ where $\lambda(h=1/2) = 2.5$ (we discuss this point in \cite{SM}).

Under non-equilibrium conditions we have also investigated the critical exponent $\nu$, defined by $\xi \propto (h-h_c)^{-\nu}$ at fixed $m_{L,R}$ \cite{SM}. Our numerical data indicate $\nu=\lambda=1/2$, which differs from the equilibrium value, $\nu=1$. 

\paragraph{Entanglement entropy:}
We now turn to the entropy content of the non-equilibrium state. The entropy of a subsystem $S_\ell$, here taken to be a segment of the chain of length $\ell$, is given by  
$E_{\t S_\ell}=-\text{Tr}\left[\hat{\rho}_{S_\ell}\ln\left(\hat{\rho}_{S_\ell}\right)\right]$, with $\hat{\rho}_{S_\ell}$ the reduced density matrix.
As the spin system can be mapped to non-interacting fermions,
the entropy can be calculated from the fermionic model \cite{Vidal.03}
and is given by
$E_{\t S_\ell}=-\text{Tr} \left [ \chi_{\t S_\ell}\ln\chi_{\t S_\ell} \right ] $, where $\chi_{\t S_\ell}$ is the single-particle density matrix restricted to $\text{S}_\ell$. 
In the limit $\ell\to\infty$, the entropy behaves as \cite{Its.2009}
\begin{equation}
E_{\t S_\ell}=l_{0}\ell+c_{0}\log\left(\ell\right)+c_{1}.
\label{eq:entanglement entropy scaling law}
\end{equation}

Ground states of gapped systems in equilibrium obey the area law,
i.e. $l_{0}=c_{0}=0$, while gapless fermions and spin chains show a universal logarithmic violation of the area law with $c_{0}=1/3$
\cite{Vidal.03, Calabrese.04}. This result is a consequence of  the violation of the area law in 1+1 conformal theories in which case $c_{0}=c/3$, where $c$ is the central charge. 
For a non-equilibrium Fermi-gas, it was shown that both
$l_{0}$ and $c_{0}$ can be non-zero \cite{Eisler.14, Ribeiro.17} and that $c_{0}$ depends on the system-reservoir coupling and is a non-analytic function of the bias \cite{Ribeiro.17}.  

For the present case the  linear coefficient $l_{0}$ is shown in Fig. \ref{fig:currentEntropy}-(a) for all phases, the details of the calculation are given in SM. 
We find that  $l_{0}$
does not vary with $m_l~ (l=L,R)$ away from the conducting phase,
depending only on the values of $h$ and $\Gamma_{l}$ (not shown in the figure). 
Moreover, $l_{0}$ vanishes within the ordered phase.
The coefficient $c_{0}$ is depicted in Fig. \ref{fig:currentEntropy}-(b). It was extracted from the mutual information, ${\cal I}\left(A,B\right)\equiv E\left(\hat{\rho}_{A}\right)+E\left(\hat{\rho}_{B}\right)-E\left(\hat{\rho}_{A+B}\right)$, of two adjacent segments $A$ and $B$ of total size $\ell$, and using that ${\cal I}\left(A,B\right)\simeq c_{0}\left[2\log\left(\ell\right)-\log\left(2\ell\right)\right]$.
We find that $c_{0}$ is non-zero in the C phase and vanishes otherwise. As in the
case of a Fermi gas, $c_{0}$ depends on the strength of the reservoir-system couplings.
In the present case, we find that it also depends on the bias potentials
away from $m_{\t L}=m_{\t R}=0$.

\begin{figure}[t]
\includegraphics[width=0.99\columnwidth]{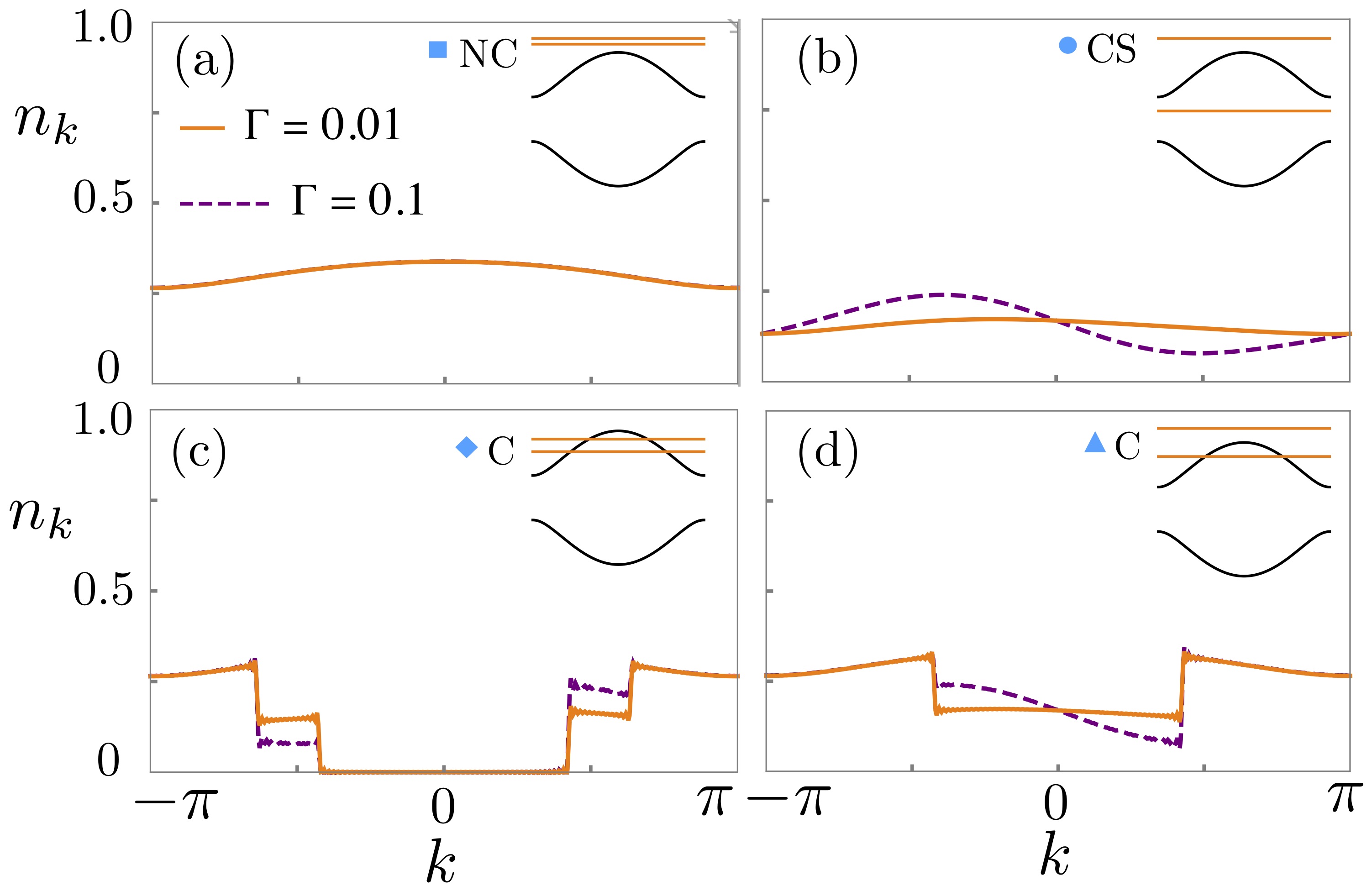}
\caption{
Distribution of excitations with momentum $k$, $n_k$, computed for different values of the reservoir-chain couplings
($\Gamma_L=\Gamma_R\equiv \Gamma$). For each panel the blue geometric symbols specifies the values of $m_\text{L}$ and $m_\text{R}$ trough  Fig.\ref{fig: phaseDiagram}-(c).    
The insets depict the energy band structure of the isolated chain (black lines), compared with
the reservoirs magnetization potential (orange lines).}
\label{fig: occupation number}
\end{figure}

\paragraph{Excitation numbers:}
In order to conceptualize  these results we  turn
to the fermionic representation.  In the infinite-volume limit, 
$L\to\infty$,  boundary effects vanish and the state becomes translationally invariant.
The Hamiltonian of the translationally
invariant chain in its diagonal representation is given by, $H=\sum_{k}\varepsilon_{k} (\gamma_{k}^{\dagger}\gamma_{k}
-1/2)$, where the operators $(\gamma_{k},\gamma_{-k}^{\dagger})^{T}=e^{i\theta_{k}\sigma_{x}}(c_{k},c_{-k}^{\dagger})^{T}$
describe the Bogoliubov excitations, $\sin\left(2\theta_{k}\right)=-2 J\sin(k)/\varepsilon_{k}$
and $\varepsilon_{k}=2\sqrt{\left(h+J\cos k\right)^{2}+\left(J\sin k\right)^{2}}$.
The excitation number $n_{k}\equiv\langle \gamma_{k}^{\dagger}\gamma_{k}\rangle _{S_\ell}$
within $S_\ell$ can be obtained from the single-particle density matrix, $\bs\chi$, numerically computed at sufficiently large $\ell$. The results are shown in Fig.\ref{fig: occupation number}, where 
the parameters used are labelled by the symbols marked in Fig. \ref{fig: phaseDiagram}-(c).
Additional distributions of $n_{k}$ are given in the SM.

For the isolated chain, the ground state is characterized by $\gamma_{k}^{\dagger}\gamma_{k}\left|\text{GS}\right\rangle =0$, {\itshape i.e.} $n_{k}=0$ for all $k$. 
In the open setup, $n_{k}=0$ also within the ordered phase, O. 
All other phases are characterized by  non-zero distributions of excitations, {\itshape i.e.} $n_{k}\neq0$. 
For the CS phases $n_{k}$ is a continuous
function of $k$ while in the C phase it may have two or four discontinuities depending on whether   
one or both of the magnetic potentials $m_L/m_R$ are located within the bands $\pm\varepsilon_{k}$, see Figs. \ref{fig: occupation number}-(c) and (d), and their insets.

Note that $n_k$ is asymmetric upon  $k\to-k$ in all conducting phases as required to maintain a net energy flow through the chain, since $\varepsilon(k)=\varepsilon(-k)$. 
In Fig.~\ref{fig: occupation number} we illustrate this feature by using a larger value of the hybridization energy, that
allows for a larger energy current thus leads to a more
asymmetric $n_k$ (see the dashed curves).

For a translational invariant system, the entanglement entropy can be obtained using the large-$\ell$ asymptotics for the determinant of Toplitz matrices, see Ref. \cite{Its.2009}.
If $n_k$ is discontinuous, the Fisher-Hartwing conjecture has to be employed.  Following the steps of Ref. \cite{Its.2009}, one concludes that $n_k\neq 0,1$ results in an extensive contribution to the entanglement entropy while every discontinuity of $n_k$ results in a logarithmic contribution to area law violation. 
This explains why $c_0\neq 0$ only within the C phase.

\paragraph{Discussion:}
We study a spin chain that can order magnetically, driven out of equilibrium by keeping the  magnetization at the two ends of the chain fixed at different values. A set of non-equilibrium phases is observed and characterized according to the conductance and the scaling of the entanglement entropy. 
This model offers a remarkable example of an extended, strongly-interacting system that can be continuously tuned from equilibrium
to non-equilibrium conditions 
and  admits an exact solution through the generalization of the Jordan-Wigner mapping. 
Moreover, we demonstrated that 
upon increasing the reservoir magnetization  a discontinuous jump of the magnetic order parameter occurs that
coincides with a divergence of the correlation length. At equilibrium, the first observation is a signature of a first-order transition, while the second is a hallmark of  continuous transitions. While this seems reminiscent of the situation that can occur at the lower critical dimension and which has been discussed in long-ranged spin chains in the context of mixed-order transitions \cite{Thouless.69, Cardy.80, Amir.14}, there are notable differences.
In the present case, the interaction is short-ranged and, more importantly, a second-order phase transition is recovered at equilibrium.  
Thus, our findings exemplify that out-of-equilibrium conditions allow for novel critical phenomena which are not possible in equilibrium. 
This kind of phase transition also differs from those  obtained for systems where dissipation is present in the bulk  which induces a change of the  dynamical critical exponent \cite{Mitra.06,Takei.10,Takei.08}. Therefore, to our best knowledge, this transition belongs to a novel universality class for which an effective field theoretic description out of equilibrium is yet to be developed. The exactly solvable model presented here should prove useful in developing such a description which will elucidate the role of interactions, {\itshape e.g.}, the presence of magnetization gradients across the chain.

From the point of view of 1D fermionic systems, 
the peculiar critical properties discussed here might provide alternative signatures of the topological transition. To address this question, it would be interesting to extend our study of criticality under nonequilibrium condition to concrete setups of semiconductor nanowires \cite{Lutchyn2010,Oreg2010,Mourik2012}.

\begin{acknowledgments}
We gratefully acknowledge helpful discussions with V.R. Vieira.
T.O. Puel acknowledges support by the NSFC (Grants No. 11750110429 and No. U1530401).
P. Ribeiro acknowledges support by FCT through the Investigador FCT contract IF/00347/2014 and Grant No. UID/CTM/04540/2013. 
S. Kirchner acknowledges support by  the National Science Foundation of China, grant No. 11774307 and the National Key R\&D Program of the MOST of China, Grant No. 2016YFA0300202. 
S. Chesi acknowledges support from NSFC (Grants No. 11574025 and No. 11750110428).
\end{acknowledgments}

\bibliographystyle{apsrev4-1}
\bibliography{refs}

\begin{thebibliography}{64}
\bibitem [{com()}]{comment1}%
  \BibitemOpen
  \href@noop {} {}\bibinfo {note} {Equation~(\ref{eq:Hamiltonian}) gives a
  superconducting gap equal to the hopping amplitude. A general Kitaev chain is
  obtained considering an anisotropic XY interaction, which induces a richer
  non-equilibrium behavior \cite{in_prep}}\BibitemShut {NoStop}%
\end{thebibliography}


\appendix

\section{A - Details of the derivations \label{sec:Notation}}

\subsection{(a) Current}

For a quadratic fermionic model the density matrix within a subsystem
S can be written as $\rho_{\t S}=\frac{e^{\Omega_{\t S}}}{\tr e^{\Omega_{\t S}}}$
where $\Omega_{\t S}=\frac{1}{2}\Psi^{\dagger}\bs{\Omega}_{\t S}\Psi$
is quadratic in the fermionic fields, with $\bs{\Omega}_{\t S}$ a
$2V_{\t S}\times2V_{\t S}$ matrix respecting the particle-hole symmetry
conditions, and where $V_{\t S}$ is the number of sites of S. In
terms of $\bs{\Omega}_{\t S}$, the single-particle
matrix $\bs{\chi}=\av{\Psi\Psi^{\dagger}}$ is given by 
\begin{align}
\bs{\chi} & =\frac{1}{1+e^{\bs{\Omega}_{\t S}}}.\label{eq:chi-1}
\end{align}
The mean value of an observable $O=\frac{1}{2}\Psi^{\dagger}\bs O\Psi$
of S, quadratic in $\Psi$ and defined by the hermitian, particle-hole symmetric matrix $\bs O$, can be obtained as 

\begin{align}
\av O=\tr\left[\rho_{\t S}\frac{1}{2}\Psi^{\dagger}\bs O\Psi\right] & =-\frac{1}{2}\tr\left[\bs O\bs{\chi}\right].
\end{align}
The expression for the energy current in the main text is obtained in this way. 

\subsection{(b) Green's function}

In the fermionic representation,
any correlation function of the chain  can be described in terms of the retarded and Keldysh components of the single-particle Green's function:
\begin{align}
\bs G^{R}\left(t,t'\right)&=-i\Theta\left(t-t'\right)\av{\left\{ \boldsymbol{\Psi}\left(t\right),\boldsymbol{\Psi}^{\dagger}\left(t'\right)\right\} }, \\
\bs G^{K}\left(t,t'\right)&=-i\av{\left[\boldsymbol{\Psi}\left(t\right),\boldsymbol{\Psi}^{\dagger}\left(t'\right)\right]}.
\end{align}
In the steady-state, the Dyson equation becomes
\begin{align}
 \bs G^{R}\left(\omega\right)&=\left[\omega-\bs H_{\text{C}}-\bs{\Sigma}^{R}\left(\omega\right)\right]^{-1},\\
\bs G^{K}\left(\omega\right)&=\bs G^{R}\left(\omega\right)\bs{\Sigma}^{K}\left(\omega\right)\bs G^{A}\left(\omega\right), 
\end{align}
with $\bs G^{A}\left(\omega\right)=\bs G^{R\dagger}\left(\omega\right)$
and where the self-energies $\bs{\Sigma}^{R/K}\left(\omega\right)=\bs{\Sigma}_{\text{L}}^{R/K}\left(\omega\right)+\bs{\Sigma}_{\text{R}}^{R/K}\left(\omega\right)$
are imposed by the reservoirs.  For the reservoirs,  
\begin{align}
\bs{\Sigma}_{l}^{K} (\omega) & = \left[\bs{\Sigma}_{l}^{R}\left(\omega\right) - \bs{\Sigma}_{l}^{A}\left(\omega\right) \right] \left[1-2 n_{\text{F},l}(\omega)\right], 
\end{align}
holds with $n_{\text{F},l}(\omega)= 1/\left( e^{\beta_l (\omega-\mu_l)} +1\right)$ being the Fermi-function, which is a manifestation of the equilibrium fluctuation dissipation relation for reservoir $l$. 
We make the simplifying
assumption that the bandwidth of the reservoirs, $J_{l}$, is much
larger than all other energy scales. In this limit \begin{align}
\bs{\Sigma}_{l}^{R}\left(\omega\right)=-i\left(\bs{\gamma}_{l}+\hat{\bs{\gamma}}_{l}\right),
\end{align}
becomes frequency
independent.  Here, $\bs{\gamma}_{l}=\Gamma_{l}\ket{r_{l}}\bra{r_{l}}$
and $\hat{\bs{\gamma}}_{l}=\Gamma_{l}\ket{\hat{r}_{l}}\bra{\hat{r}_{l}}$,
and where $\ket r$ and $\ket{\hat{r}}=\bs S\ket r$ are single-particle
and hole states. $\Gamma_{l}=\pi J_{l}^{'2}D_{l}$ is the hybridization
energy scale and $D_{l}$ is the local density of states of reservoir $l$.
The non-Hermitian
single-particle operator
\begin{align}
\bs K=\bs H_{\t C}-i\sum_{l=\text{L,R}}\left(\bs{\gamma}_{l}+\hat{\bs{\gamma}}_{l}\right),
\end{align}
introduced in the main text, possesses
eigenvalues $\lambda_{\alpha}$ and corresponding right and left
eigenvectors $\ket{\alpha}$ and $\bra{\tilde{\alpha}}$, in terms of  which
the Green function $\bs G^{R}$ is simply given by \begin{align}
\bs G^{R}\left(\omega\right)=\sum_{\alpha}\ket{\alpha}\left(\omega-\lambda_{\alpha}\right)^{-1}\bra{\tilde{\alpha}}.
\end{align}
Equal-time observables can thus be obtained from the single-particle density
matrix, defined as 
\begin{align}
\boldsymbol{\chi}& =\frac{1}{2}\left[i\int\frac{d\omega}{2\pi}\boldsymbol{G}^{K}\left(\omega\right)+\bs 1\right].
\end{align}
The explicit evaluation of this expression yields Eq.~(2) of the main text.

\subsection{(c) Two-spin correlation function}

By symmetry arguments, for finite $L$, $\av{\sigma_{m}^{x}}=0$.
Thus, the $\av{\sigma_{m}^{x}\sigma_{n}^{x}}$ correlation function, for $m$ and $n$ (with $m>n$) belonging
to a subsystem S, can be written as 
\begin{multline}
C_{mn}^{xx}=\av{\sigma_{m}^{x}\sigma_{n}^{x}}\\
=\tr\left[e^{i\pi\sum_{j=n}^{m}c_{j}^{\dagger}c_{j}}\left(-c_{m}^{\dagger}+c_{m}\right)\left(c_{n}^{\dagger}+c_{n}\right)\rho_{\t S}\right].\label{eq:c_xx}
\end{multline}
We now re-write Eq.~(\ref{eq:c_xx}) in terms of the operators $\Omega_{1}=\frac{1}{2}\Psi^{\dagger}\bs{\Omega}_{1}\Psi$,
with $\bs{\Omega}_{1}=
i\pi\sum_{j=n}^{m}\left(\ket j\bra j-\ket{\hat{j}}\bra{\hat{j}}\right)
$, $A=\left(-c_{m}^{\dagger}+c_{m}\right)\left(c_{n}^{\dagger}+c_{n}\right)=\frac{1}{2}\Psi^{\dagger}\bs A\Psi$,
with 
\begin{align}
\bs A & =-2\left[\ket{m_{-}}\bra{n_{+}}+\ket{n_{+}}\bra{m_{-}}\right],\label{eq:A}
\end{align}
 $\ket{r_{\pm}}=\left(\ket r\pm\ket{\hat{r}}\right)/\sqrt{2},$ and
$\rho_{\t S}$ as in previous section. These definitions lead to:
\begin{align}
C_{mn}^{xx} & =-ie^{\frac{i}{2}\pi\left(m-n+1\right)}T,
\end{align}
 with 
\begin{align}
T & =\frac{\tr\left[e^{\Omega_{1}}e^{i\frac{\pi}{2}A}e^{\Omega_{\t S}}\right]}{\tr e^{\Omega_{\t S}}},
\end{align}
using that $e^{i\frac{\pi}{2}A}=iA$.

\begin{figure*}
\centering{}\includegraphics[width=0.95\linewidth]{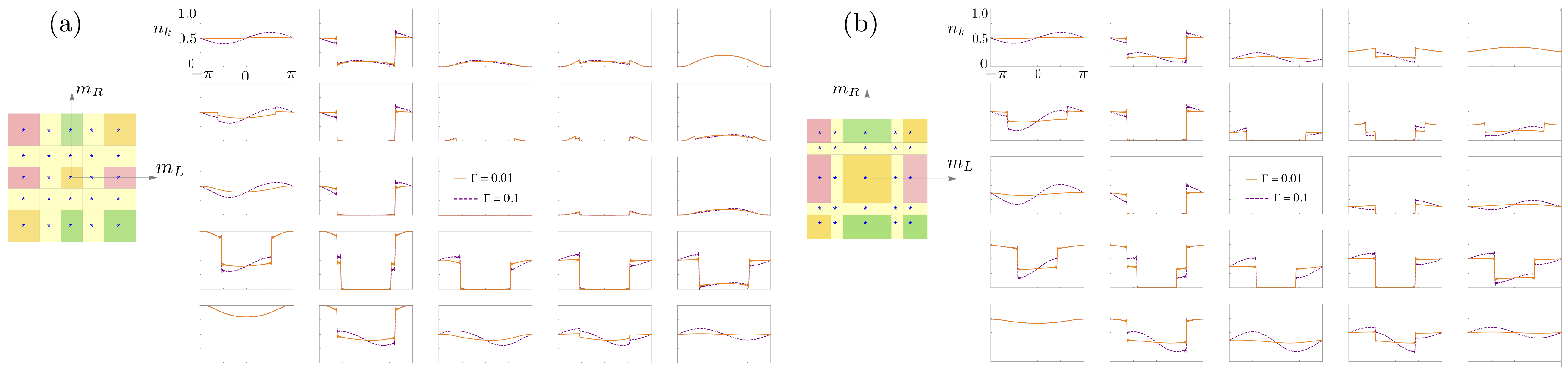}
\caption{Panel (a) shows the excitation numbers $n_{k}$ in each phase of the phase diagram at $h=0.2$, while panel (b) shows the excitation numbers at the special point $h=1/2$.}
\label{fig:all excitations}
\end{figure*}

We now use the Levitov-Lesovik formula \cite{Levitov1993,Levitov1996} to evaluate the trace,
\begin{align}
T= & \frac{\det\left[ 1+e^{\bs{\Omega}_{1}}e^{i\frac{\pi}{2}\bs A}e^{\bs{\Omega}_{\t S}} \right]^{\frac{1}{2}}}{\det\left[ 1+e^{\bs{\Omega}_{\t S}} \right]^{\frac{1}{2}}},
\end{align}
and ~\eqref{eq:chi-1}, to write 
\begin{align}
T & =\det\left[1+\left(1-\bs{\chi}\right)\left(-1+e^{\bs{\Omega}_{1}}e^{i\frac{\pi}{2}\bs A}\right)\right]^{\frac{1}{2}}.
\end{align}
This expression can be further simplified noting that $e^{\bs{\Omega}_{1}}=\bar{\bs P}-\bs P$
with 
\begin{align*}
\bs P & =\sum_{j=n}^{m}\left[\ket i\bra i+\ket{\hat{i}}\bra{\hat{i}}\right]
\end{align*}
and $\bar{\bs P}=1-\bs P$. Since $\bs A\bs P=\bs P\bs A=\bs A$ we
can write 
\begin{align}
T & =\det\left[\bs{\chi}+\left(1-\bs{\chi}\right)\left(\bar{\bs P}-\bs Pe^{i\frac{\pi}{2}\bs A}\right)\right]^{\frac{1}{2}}. \label{S8}
\end{align}
This expression can be simplified noting that, since $\bar{\bs P}\left[\bs{\chi}+\left(1-\bs{\chi}\right)\left(\bar{\bs P}-\bs Pe^{i\frac{\pi}{2}\bs A}\right)\right]\bar{\bs P}=\bar{\bs P}$
and $\bs P\left[\bs{\chi}+\left(1-\bs{\chi}\right)\left(\bar{\bs P}-\bs Pe^{i\frac{\pi}{2}\bs A}\right)\right]\bar{\bs P}=0$,
the determinant is solely determined by the projection onto the
subspace where $\bs P$ acts as the identity. We define the restriction
of $\bs{\chi}$ and $\bs A$ to that subspace, spanned by the
sites $n\le r\le m$, as
\begin{align}
\tilde{\bs{\chi}} & =\bs p^{T}\bs{\chi}\bs p\\
\tilde{\bs A} & =\bs p^{T}\bs A\bs p
\end{align}
where $\bs P=\bs p\bs p^{T}$and $\bs p^{T}\bs p=\bs 1$. We can now write 
\begin{align}
T & =\det\left[\tilde{\bs{\chi}}\left(e^{i\frac{\pi}{2}\tilde{\bs A}}+1\right)-1\right]^{\frac{1}{2}}.
\end{align}
We further note that 
\begin{align}
e^{i\frac{\pi}{2}\tilde{\bs A}}-1 & =-2 \tilde{\bs{Q}}  
\end{align}
with $\tilde{\bs{Q}}=\ket{n_{+}}\bra{n_{+}}+\ket{m_{-}}\bra{m_{-}}$, thus
\begin{align}
T & =\det\left[2\tilde{\bs{\chi}}\left(1-\tilde{\bs{Q}}\right)-1\right]^{\frac{1}{2}}.
\end{align}
Again, this expression can be simplified in a way similar to Eq.~(\ref{S8}) noting that $\left(1-\tilde{\bs{Q}}\right)\left[2\tilde{\bs{\chi}}\left(1-\tilde{\bs{Q}}\right)-1\right]\tilde{\bs{Q}}=0$
and $\tilde{\bs{Q}}\left[2\tilde{\bs{\chi}}\left(1-\tilde{\bs{Q}}\right)-1\right]\tilde{\bs{Q}}=-\tilde{\bs{Q}}$.
Thus, we can define $\mathbb{P}_{mn}=\bs P\left(1- \bs{Q}\right)$, where $\bs{Q}$ is the extension of $\tilde{\bs{Q}}$ to the entire space. The explicit expression of $\mathbb{P}_{mn}$ is
given in the main text. 
For $\bar{\bs q}$ such that $\mathbb{P}_{mn}=\bar{\bs q}\bar{\bs q}^{T}$
and $\bar{\bs q}^{T}\bar{\bs q}=\bs 1$, we obtain 
\begin{align}
T & =\det\left[2\bar{\bs q}^{T}\bs{\chi}\bar{\bs q}-1\right]^{\frac{1}{2}}
\end{align}
and recover the expression 
\begin{align}
C_{mn}^{xx} & =\det\left[i\left(2\bar{\bs q}^{T}\bs{\chi}\bar{\bs q}-1\right)\right]^{\frac{1}{2}}
\end{align}
given in the main text. $C_{mn}^{yy}$ can be obtained
in a similar fashion.

\section{B - Additional numerical results}

For completeness, the following provides a complementary set of numerical results to those given in the main text.

\subsection{(a) Excitations numbers}

Fig. \ref{fig:all excitations}-(a) illustrates the excitation numbers $n_{k}$ in all  regions of the phase diagram, 
for the same set of parameters used in the main text: $J=1$, $h=0.2$, $\Gamma_{\text R} = \Gamma_{\text L} = 0.01$ or $0.1$, and zero temperature.
This choice of parameters yields $m_1 = 2(-h+1)=1.6$ and $m_2 = 2(h+1) = 2.4$.
We have used $L=500$ for which finite size effects are negligible. 

As noted in the main text, the asymmetry upon changing $k \rightarrow -k$ of the conducting phases is enhanced by a larger value of the hybridization between the chain and the reservoirs. 
The panels of $n_k$ follow the same order as the markers depicted in the phase diagram.

\begin{figure}
\centering{}\includegraphics[width=0.95\columnwidth]{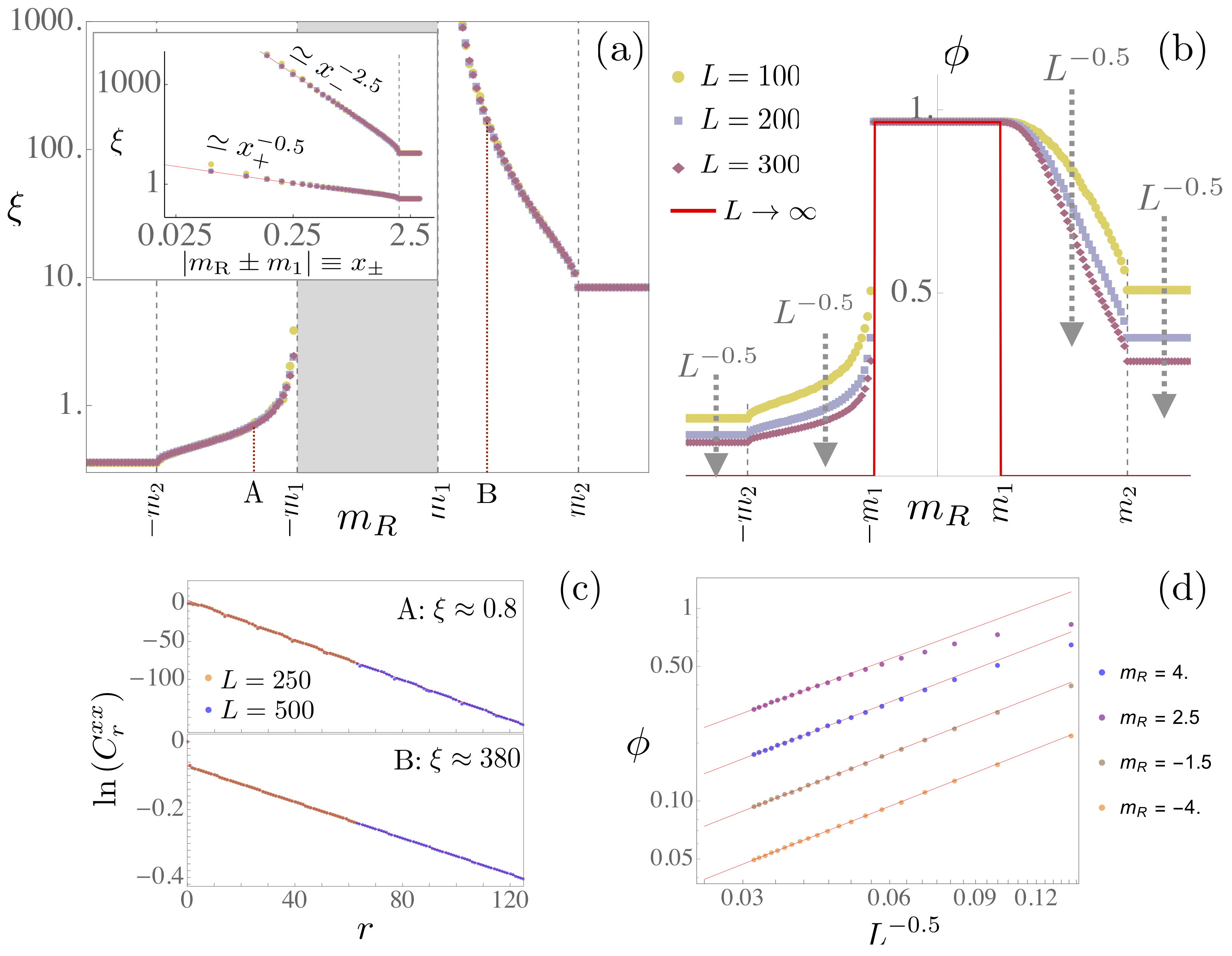}
\caption{Correlation length (a) and order parameter (b) for $h=1/2$. The inset shows the scaling of $\xi$ near the transition at $m_R = \pm m_1$. The fittings to compute $\xi$ are exemplified in (c) for two points (A and B) marked in the first panel. The finite size scaling behaviour of the order parameter is shown in (d), for four points inside different disordered phases (marked by arrows in panel (b)).}
\label{fig:correlation length h0.5}
\end{figure}

\section{(b) Case $h=1/2$ \label{sec:h=1/2}}

Here we expand on the special case of $h=1/2$, briefly mentioned in the main text, which leads to a different universality class, {\itshape i.e.}, to different critical exponents.
Figs. \ref{fig:correlation length h0.5}-(a) and (b) show the correlation length and order parameter, respectively. For both panels $m_{\text L} = 0.5$ was used.
The inset shows the correlation length diverging as $\xi \propto \left|m_{\text R} \pm m_1 \right|^{-\nu}$ for $m_{\text R} \rightarrow \mp m_1$.
Note that $\nu=1/2$ for $m_{\text R} \rightarrow -m_1$, while $\nu=5/2$ for $m_{\text R} \rightarrow +m_1$.
Typical fittings of the correlation function $C^{xx}_r$, computed to obtain the correlation length, are illustrated in Fig. \ref{fig:correlation length h0.5}-(c).
Note that for this choice of magnetic field $h$ one obtains $m_1 = 1$ and $m_2 = 3$.
Finally, a finite size scaling analysis of the order parameter is shown in Fig. \ref{fig:correlation length h0.5}-(d) for $L \in [10^2 , 10^3]$.

The special behavior for $h=1/2$  can be understood by analyzing its excitation numbers.
Fig. \ref{fig:all excitations}-(b) illustrates $n_{k}$ in all  regions of the phase diagram, under the same conditions of Fig. \ref{fig:all excitations}-(a).
The difference appears on values of $m_{\text R}$ and $m_{\text L}$ for which the excitations raise continuously from zero, as we drive the system out of the ordered phase.
Note, in fact, that when $m_R \rightarrow m_1$ the disordered phase is characterized by $n_{k=\pm\pi} =0$, which corresponds to the anomalous exponent $\nu = 5/2$. On the other hand, at the $m_R \rightarrow -m_1$ phase boundary one has $n_{k=\pm\pi} \neq 0$, giving the same exponent $\nu=1/2$ discussed in the main text.

\section{(c) Critical exponent $\nu$ far from equilibrium \label{sec:nu}}

\begin{figure}[H]
\centering{}\includegraphics[width=0.95\columnwidth]{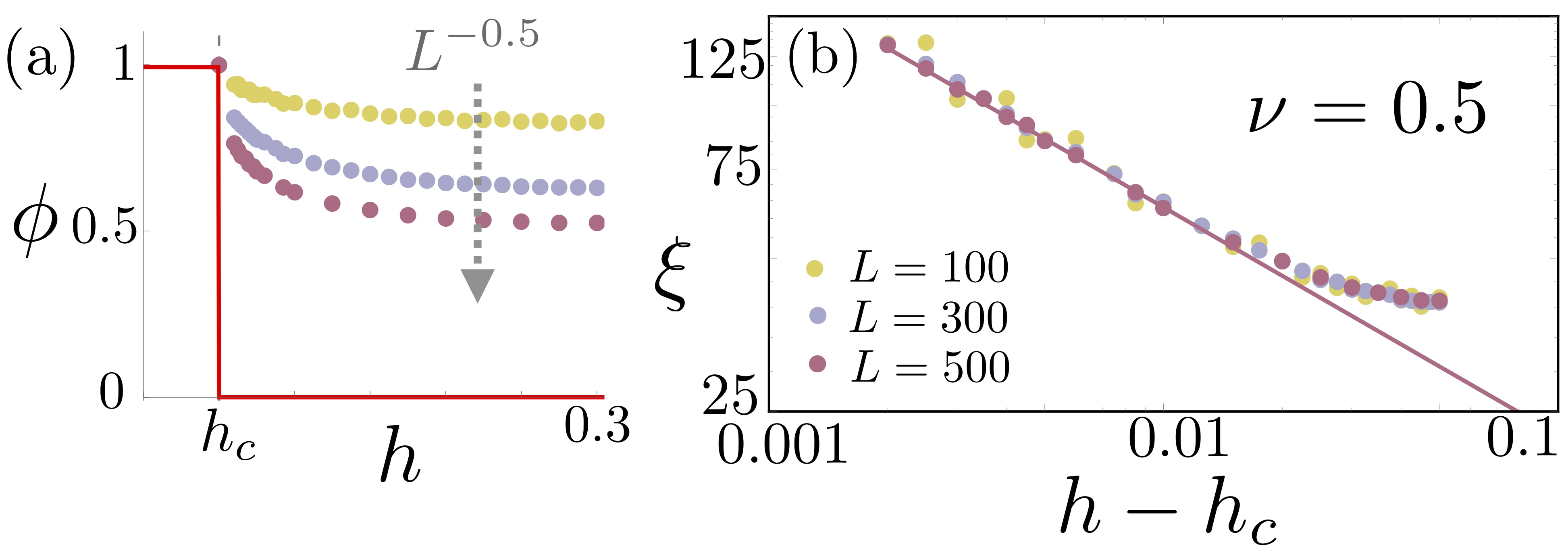}
\caption{Panel (a) shows the order parameter $\phi$ for a non-equilibrium steady-state transition induced by $h$. The solid red line indicates the thermodynamic limit $L \rightarrow \infty$. Panel (b) shows the correlation length $\xi$ near to the critical point $h_c$ as we change $h$.}
\label{fig:nu}
\end{figure}

Here we present results for the order-disorder non-equilibrium phase transition induced by the transverse magnetic field $h$.
Fig.\ref{fig:nu}-(a) shows the order parameter $\phi$ near the critical point $h_c = 0.25$ obtained for $m_{\text{L}}=0$ and $m_{\text{R}}=1.5$.
Fig.\ref{fig:nu}-(b) shows the correlation length $\xi$ near the critical point $h_c$. 
The associated critical exponent $\nu\simeq 0.5$ is extracted from the correlation length $\xi$ by fitting to a power-law dependence $\xi \propto \left( h - h_c \right)^{-\nu}$. 
These results show that a first order transition with essentially the same features as the one shown in the main text can also be assessed through varying $h$, rather then $m_\text{R}$ or $m_\text{L}$, with the same critical exponents $\nu=\lambda=1/2$.

\end{document}